\documentclass[10pt,a4paper]{article}

\title{Evanescent gravitational mass\\
{\small Physics without geometry}}

\author{Llu\'{\i}s\ Bel\\
\emph{wtpbedil@lg.ehu.es}
}
\begin{document}

\maketitle

\begin{abstract}

A simple and {\it innocent} modification of Poisson's equation leads to a modified Newtonnian theory of gravitation where
a localized and {\it positive} energy density of the gravitational field contributes to its own source. The result is
that the total {\it active gravitational mass} of a compact object is the sum of its {\it proper mass} and an {\it evanescent gravitational mass} which is a mass equivalent to the gravitational energy.

\end{abstract}

Conventional wisdom subordinates the concept of the energy of the Newtonnian gravitational field to the potential energy of the source. As if the unique interest of the first was to provide us with an alternate method to calculate the second. We are thus led to accept that the energy of the gravitational field is {\it negative}.

Since the potential function $V$ is defined only up to an additive constant, conventional wisdom has also made us comfortable with choosing this constant such that $V_\infty$ is zero, while identifying it with the square of a velocity opens new opportunities of interpretation.

The first section below introduces the notations, develops somewhat the preceding remarks and serves as a transition to the second section. There we define the energy density $\sigma_g$ of the gravitational field as a localized {\it positive} quantity, unrelated {\it a priori} to the potential energy density, and modify the Poisson equation The new equation is not linear and is formally equivalent to the Poisson equation with a source term that is the sum of the {\it proper mass density} $\rho_p$ and the {\it evanescent gravitational mass} one $\rho_g=\sigma_g/c_g^2$, where $c_g$, is a new parameter with the physical dimensions of a velocity, that may be related to $V_\infty$ or not.

An obvious consequence of this modified theory is that the active gravitational mass is always {\it greater}  than its proper mass. The {\it gain factor} might have any value greater than 1, depending on the object under consideration.

The last section contents a few elements of comparison between the plain Newtonnian theory and the modified one, as well as some comments about possible choices of the new parameter $c_g$. Its value should be derived from an appropriate confrontation of the new theory with observational data, an undertaking that is well beyond the scope of this paper and the competence of its author.

{\it Newtonnian gravity}.- Newtonnian gravity is based on Poisson's equation:

\begin{equation}
\label{1.1}
\triangle V=4\pi G\rho_p
\end{equation}
This equation being linear the gravitational potential $V$ is defined only up to an additive constant that it is usually chosen so that V tends to zero at infinity. We shall see in a moment that making a different choice opens interesting new perspectives.

We shall restrict the scope of this paper to the case where the source of the field is a spherical body with finite radius $R$ and proper mass density $\rho_p$ depending only on the distance $r$ at the center. Then we have:

\begin{equation}
\label{1.3}
\frac{d^2V}{dr^2}+\frac2r\frac{dV}{dr}=4\pi G\rho_p
\end{equation}
Setting:

\begin{equation}
\label{1.4}
W=r^2\frac{dV}{dr}
\end{equation}
Equ. (\ref{1.3}) is equivalent to the couple of equations:

\begin{equation}
\label{1.5.1}
\frac{dW}{dr}=4\pi G\rho_p r^2,  \ \ (a)\ \ \
\frac{dV}{dr}=\frac{W}{r^2} \ \ (b)
\end{equation}
whose solution is, requiring the value of $W$ at the center to be zero:

\begin{equation}
\label{1.6.1}
W=4\pi G\int_0^r\rho_p r^2\,dr, \ \ (a)\ \ \
V=-\int_r^\infty\frac{W}{r^2}\,dr+V_\infty \ \ (b)
\end{equation}
where $V_\infty$ is an arbitrary constant. Let us call {\it proper mass} of the source the quantity $M_p$:

\begin{equation}
\label{1.6.3}
M_p=4\pi\int_0^\infty\rho_p r^2\,dr,
\end{equation}
and {\it active gravitational mass} the quantity $M_a$ in the formula:

\begin{equation}
\label{1.6.4}
GM_a=\lim_{r->\infty}W
\end{equation}
$M_p$ and $M_a$ are different physical quantities, but it follows from (\ref{1.6.1}a) that numerically we have $M_a=M_p$.

We define the energy density of the gravitational field as the {\it positive} quantity:

\begin{equation}
\label{1.8}
\sigma_g=\frac{1}{8\pi G}\frac{W^2}{r^4}
\end{equation}
and using (\ref{1.1}) and integrating by parts  the integral giving the total energy we obtain:

\begin{equation}
\label{1.9}
E_g=4\pi\int_0^\infty\sigma_g r^2\,dr=\frac{1}{2 G}(VW)_\infty-2\pi\int_0^R \rho_pVr^2\,dr
\end{equation}
If we choose to set $V_\infty=0$ then we have $E_g=-U_p/2$, where $U_p$ is the gravitational potential energy\,\footnote{See for instance \cite{Chandra} III.2}:

\begin{equation}
\label{1.9.1}
U_p=4\pi\int_0^R V\rho_p r^2\,dr
\end{equation}
But otherwise we can write the preceding formula (\ref{1.9}) as:

\begin{equation}
\label{1.10}
M_a=\frac{1}{V_\infty}(2 E_g +U_p)
\end{equation}
A second natural choice would be to set:

\begin{equation}
\label{1.10.1}
V_\infty=\int_0^\infty \frac{W}{r^2}\,dr
\end{equation}
in which case one has $V_0=0$ at the center and $V$ positive all the way for $r>0$.

We shall assume that $V_\infty$ is positive and write $V_\infty=c_g^2$, so that the quantity:

\begin{equation}
\label{1.10.2}
\rho_g=\frac{\sigma_g}{c_g^2}
\end{equation}
will be a mass density equivalent to the gravitational energy density.

One of the potential benefits from choosing $V_\infty$ satisfying (\ref{1.10.1}) is that the {\it  gravitational  active mass-function}:

\begin{equation}
\label{1.10.3}
M_a(r)=4\pi\int_0^r(2\rho_g+(V/c_g^2)\rho_p)r^2\,dr
\end{equation}
is then a positive monotonously increasing function of $r$ tending to $M_a$ as $r$ goes to $\infty$.

The formula (\ref{1.10}) would be a potentially interesting one if we could relate $U_p$ to some {\it luminosity mass}. But since this is doubtful we proceed to the next section.

%%%%%%%%%%%%%%%%%%%%%%%%%%%%%%%%%%%%%%%%%%%%%%%%%%%%%%%%

\vspace{1cm}

{\it Ponderous gravitational energy}.- In this section  we consider the following modification of Eq. (\ref{1.1}):

\begin{equation}
\label{2.1}
\triangle V=4\pi G(\rho_p+\rho_g)
\end{equation}
where we maintain the definitions of $\sigma_g$ and $\rho_g$ in (\ref{1.8}) and (\ref{1.10.2}). We shall maintain also the definitions of $M_p$ in (\ref{1.6.3}) and $M_a$ in (\ref{1.6.4}), while the discussion about the plausible values of the parameter $c_g^2$ will be made latter.

Eq. (\ref{2.1}) is equivalent to the couple of equations:

\begin{equation}
\label{2.2}
\frac{dW}{dr}=4\pi G(\rho_p+\rho_g)r^2, \quad \frac{dV}{dr}=\frac{W}{r^2}
\end{equation}
The first one is a Ricatti equation that we integrate assuming that $\rho_p$ is constant and requiring $(W_i/r^2)(0)=0$. The interior solution $W_i$, i.e. for $r<R$, is then:

\begin{equation}
\label{2.3}
W_i=2rc_g^2\left(1+\frac{\alpha r}{2}F(r)\right),
\end{equation}
where:

\begin{equation}
\label{2.3ter}
F(r)=\frac{\cos\alpha r}{1-\sin\alpha r}
+\frac{2(1+\sin\alpha r)}{\cos\alpha r(\cos\alpha r-1-\sin\alpha r)}
\end{equation}
or\,\footnote{Courtesy of J.\ M.\ Aguirregabiria and J.\ Mart\'{\i}n}:

\begin{equation}
\label{2.3qto}
F(r)=-\cot\left(\frac{\alpha}{2}r\right)
\end{equation}
and:

\begin{equation}
\label{2.3.1}
\alpha=\frac{(8\pi G\rho_p)^{1/2}}{c_g}
\end{equation}
While the exterior solution $W_e$, i.e. for $r>R$, is:

\begin{equation}
\label{2.4}
W_e=\frac{2rc_g^2}{1+2C_er}
\end{equation}
where $C_e$ is a constant of integration that has to be fixed requiring the continuity of $W$ across the boundary of the source:

\begin{equation}
\label{2.5}
W_i(R)=W_e(R)
\end{equation}
On the other hand, using the definition (\ref{1.6.4}) of $M_a$, we have:

\begin{equation}
\label{2.5.0.1}
G M_a=\frac{c_g^2}{C_e}
\end{equation}
so that setting:

\begin{equation}
\label{2.5.0.1.0}
C_e=\frac{c_g^2}{G M_p X}
\end{equation}
where $X=M_a/M_p$ is the {\it gain factor}, we can calculate this quantity using (\ref{2.3}), (\ref{2.4}) and (\ref{2.5}), obtaining after some elementary algebra:

\begin{equation}
\label{2.5.0.2}
X=-\frac{8\sqrt{3}+12\sqrt{\lambda} F(\lambda)}{3\lambda\sqrt{\lambda}F(\lambda)}
\end{equation}
where $\lambda$ is the {\it compactness parameter}:

\begin{equation}
\label{2.6}
\lambda=\frac{2 G M_p}{c_g^2R}
\end{equation}
and:

\begin{equation}
\label{2.6.1}
F(\lambda)=\frac{1+\cos\sqrt{3\lambda}-\sin\sqrt{3\lambda}}{1-\cos\sqrt{3\lambda}-\sin\sqrt{3\lambda}}
\end{equation}
$X$ is a monotonous increasing function of $\lambda$ becoming infinite for:

\begin{equation}
\label{2.8}
\lambda_{max}=\frac{\pi^2}{3}=3.289868134
\end{equation}
These are the values of $X$ for $\lambda=1,2,3$:

\begin{tabular}{|l|l|l|}
\hline
$\lambda$ & $X$\\
\hline
1 & \ 1.43\\
2 & \ 2.53\\
3 & 11.20\\
\hline
\end{tabular}

For small values of $\lambda$, $X$ can be approximated to:

\begin{equation}
\label{2.7}
X=1+\frac{3}{10}\lambda+\frac{51}{560}\lambda^2.
\end{equation}.

%%%%%%%%%%%%%%%%%%%%%%%%%%%%%%%%%%%%%%%%%%%%%%%%%%%%%%%%%%

{\it Final comments}.- To compare the two theories, the Newtonnian and the modified one, we consider a test particle moving freely in the exterior gravitational field, $r\geq R$, of the spherical body. Let $a_0$ be the  acceleration according to Newtonnian theory and let $a_1$ be that one according to the modified theory:

\begin{equation}
\label{2.7.1}
a_0=\frac{G M_p}{r^2}, \quad a_1={2 c_g^2}{r}\left(1+\frac{4}{\lambda X}\frac{r}{R}\right)^{-1}
\end{equation}.

Defining the {\it anomaly} $\epsilon$ as follows:

\begin{equation}
\label{x.1}
\epsilon=\frac{a_1}{a_0}-1
\end{equation}
we have, using the available definitions:

\begin{equation}
\label{x.2}
\epsilon=\frac{4r}{\lambda R}\left(1+\frac{4 r}{\lambda X R}\right)^{-1}-1
\end{equation}
The larger value that this anomaly $\epsilon$ can be at a distance $r$ is:

\begin{equation}
\label{x.4}
\epsilon_{max}=\frac{4}{\lambda_{max}}\frac{r}{R}-1
\end{equation}
where $\lambda_{max}$ is the value given in (\ref{2.8}).

For small values of $\lambda$, using the approximate expression (\ref{2.7}) in (\ref{x.2}) and approximating it again we get:

\begin{equation}
\label{x.3}
\epsilon=\left(\frac{3}{10}-\frac{1}{4}\frac{R}{r}\right)\lambda
\end{equation}

At this point, to push further the comparison between the two theories and eventually when trying to falsify the modified one will requires to face the problem  of choosing $c_g$. This is something that we shall not attempt to do it here. But we comment below several open possibilities:

(i) Choosing $c_g=c$ is a natural choice nowadays. It would not have been in the 19th century. With this choice two bodies with the same compactness parameter $\lambda$ but quite different {\it visible parameters} $M_p$ and $R$ will have the same gain factor $X$ and give raise to the same qualitative gravitational field. This is the case with th Sun, $\lambda=4.3\,10^{-6}$, and the huge galaxy NGC 4472, $\lambda=3.8\,10^{-6}$, which has a proper mass of the order  of $10^{12}$ solar masses and a diameter of about $50$ kpc. It looks surprising that two objects so different from the morphological point of view could share some essential features of their gravitational fields.

R being the radius of the sun let us assume that $r\approx215\times R$ which is approximately 1 AU. We have then, calculating the anomaly defined in (\ref{x.2}), $\epsilon=1.27\,10^{-6}$. For much larger values of $r$ the anomaly will remain approximately constant to $\epsilon=1.28\,10^{-6}$.

(ii) Choosing $c_g$ to be a universal constant different from $c$ would be a major change in our understanding of gravity, but we shall have to wait until observational bounds on the value of $c_g$ become available. This possibility does not cure the embarrassing precedent remark coming from the similitude of the values of $\lambda$ for the Sun and NGC 4472

(iii) considering $c_g$ as an {\it invisible phenomenological parameter} depending on the morphology of the object under consideration severely restricts the predictability of the modified theory, but it may be useful to to take into account the morphology of the source of the field and fit scandalous anomalous data.

(iv) Maintaining the definition (\ref{1.10.1}) of $V_\infty$ in the modified theory and setting $c_g^2=V_\infty$ leads to the equation:

\begin{equation}
\label{2.9}
c_g^2=\int_0^R W_i\,dr+\int_R^\infty W_e\,dr
\end{equation}
Using (\ref{2.3}) and (\ref{2.4}) $c_g^2$ drops out and calculating the integrals we get:

\begin{equation}
\label{2.10}
\ln\left(\frac{\alpha^2 R^2}{4}\frac{1+\tan^2 (\alpha R/2)}{\tan^2 (\alpha R/2)}\right)+\ln\left(\frac{1+2 C_e R}{2 C_e R}\right)^2=1
\end{equation}
that, using the available definitions of $\alpha,\ \lambda$, and $X$, gives:

\begin{equation}
\label{2.11}
\left(1+\frac{\lambda X}{4}\right)^2\frac{3 \lambda}{4\sin^2(\sqrt{3\lambda}/2)}= {\bf e}
\end{equation}
which is an equation for $\lambda$. Its unique acceptable solution is $\lambda=1.126342730$. With this choice $c_g$ remains arbitrary and $\lambda$ takes a fixed value independent of the visible parameters. The corresponding common gain factor being: $X=1.513581269$

(v) Choosing $c_g(t)$ a function of time would allow to discuss possible evolutionary  effects.

(vi) Maybe the best choice: $c_g=\infty$

\end{document}